\documentclass[DIV=calc,paper=a4,fontsize=11pt,twocolumn]{scrartcl} % KOMA-article class

\usepackage[english]{babel}
\usepackage[protrusion=true,expansion=true]{microtype}
\usepackage{amsmath,amsfonts,amsthm}
\usepackage[final]{graphicx}
\usepackage{xcolor}
\usepackage[normal,small,hypcap,up,labelfont=bf,textfont=it]{caption}
\usepackage{subfig}
\usepackage{booktabs}
\usepackage{fix-cm}
\usepackage{amssymb,amsfonts}
\usepackage{dsfont}
\usepackage{bbm}
\usepackage{cite}
\usepackage[utf8]{inputenc}
\usepackage[perpage,symbol*]{footmisc}
\usepackage[varg]{txfonts}
\usepackage{balance}
\usepackage{fancyhdr}
\PassOptionsToPackage{hyphens}{url}\usepackage[pdfencoding=auto,psdextra]{hyperref}
\usepackage{bookmark}
\usepackage{verbatim}
\usepackage{fontenc}
\usepackage{cuted}
\DeclareCaptionFont{mycolor}{\color[HTML]{000000}}
\usepackage{sectsty}													% Custom sectioning (see below)
\allsectionsfont{%															% Change font of al section commands
\color[HTML]{31ADF3}\usefont{OT1}{phv}{b}{n}%										% bch-b-n: CharterBT-Bold font
}

\sectionfont{%																% Change font of \section command
\color[HTML]{31ADF3}\usefont{OT1}{phv}{b}{n}%										% bch-b-n: CharterBT-Bold font
}

\usepackage{fancyhdr}												% Needed to define custom headers/footers
\usepackage{lettrine}
\newcommand{\initial}[1]{%
\lettrine[lines=3,lhang=0.3,nindent=0em]{
\color[HTML]{31ADF3}
{\textsf{#1}}}{}}

\usepackage{titling}															% For custom titles

\newcommand{\HorRule}{\color[HTML]{31ADF3}%			% Creating a horizontal rule
\rule{\linewidth}{1pt}%
}

\pretitle{\vspace{-30pt} \begin{flushleft} \HorRule
\fontsize{34}{34} \usefont{OT1}{phv}{b}{n} \color[HTML]{31ADF3} \selectfont
}
\title{How Do the Probabilities Arise in Quantum Measurement?}					% Title of your article goes here
\posttitle{\par\end{flushleft}\vskip 0.5em}

\preauthor{\begin{flushleft}\large \lineskip 0.5em \usefont{OT1}{phv}{b}{sl} \color[HTML]{31ADF3}}
\author{Mani L. Bhaumik\\[8pt]}											% Author names go here
\postauthor{\footnotesize \usefont{OT1}{phv}{m}{sl} \color[HTML]{000000}
Department of Physics and Astronomy, University of California, Los Angeles, USA. E-mail: \href{mailto:bhaumik@physics.ucla.edu}{bhaumik@physics.ucla.edu}\\[10pt]		% Institutions of authors
\scriptsize\usefont{OT1}{phv}{m}{n} \color[HTML]{31ADF3}{\textbf{Editors: \emph{Zvi Bern} \& \emph{Danko Georgiev}} }\\[5pt]
\color[HTML]{000000}{Article history: Submitted on December 7, 2021;  Accepted on December 15, 2021; Published on December 17, 2021.}
\par\end{flushleft}\HorRule}

\date{}																				% No date

\begin{document}
\maketitle
\thispagestyle{fancy} 			% Enabling the custom headers/footers for the first page
% The first character should be within \initial{}
\initial{A}\textbf{satisfactory resolution of the persistent quantum measurement problem remains stubbornly unresolved in spite of an overabundance of efforts of many prominent scientists over the decades. Among others, one key element is considered yet to be resolved. It comprises of where the probabilities of the measurement outcome stem from. This article attempts to provide a plausible answer to this enigma, thus eventually making progress toward a cogent solution of the longstanding measurement problem.\\ Quanta 2021; 10: 65--74.}

\begin{figure}[b!]
\rule{245 pt}{0.5 pt}\\[3pt]
\raisebox{-0.2\height}{\includegraphics[width=5mm]{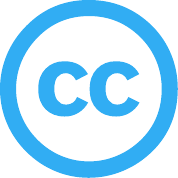}}\raisebox{-0.2\height}{\includegraphics[width=5mm]{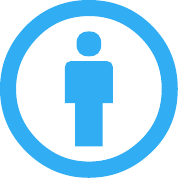}}
\footnotesize{This is an open access article distributed under the terms of the Creative Commons Attribution License \href{http://creativecommons.org/licenses/by/3.0/}{CC-BY-3.0}, which permits unrestricted use, distribution, and reproduction in any medium, provided the original author and source are credited.}
\end{figure}

\section{Introduction}

The quantum measurement problem is considered as one of the important
unresolved problems in physics although its origin goes back to the
very inception of quantum mechanics nearly a century ago. In spite
of an abundance of ideas that have been pursued throughout the decades
resulting in countless articles, absence of a satisfactory explanation
of the processes involved in the quantum to classical transition,
also known as quantum measurement problem, has tenaciously persisted
as a frustrating feature of quantum physics. This is very possibly
because it involves the most distinctive characteristics of superposition
of states in the quantum arena. A quantum state that has not yet been
measured is in a superposition of two or more possible states of definite
eigenvalue, the superposition differing qualitatively from any one
of those states. There is no apparent manifestation of superposition
in our familiar daily classical world, where tangible measurements
of the quantum states are accomplished. In fact when we try to extrapolate
the quantum superposition to classical domain in its entirety, we
end up with such absurdity as the existence of a simultaneously dead
and alive Schr\"{o}dinger's cat.

Yet it is an undeniable fact that the simultaneous existence of both
the microscopic quantum world and the macroscopic classical world
is essential for reality in a rather inseparably intertwined manner.
For example, we humans are large and as such belong to the macroscopic
classical world. However, everyone of us consists of about $7\times 10^{27}$
atoms \cite{Kross-2021} each containing additional elementary particles,
all of which are in the microscopic quantum domain. Thus, we and everything
else around us inevitably belong simultaneously both to the microscopic
quantum as well as the macroscopic classical domain of the universe
without ever paying much attention to this momentous reality. Indubitably
the quantum realm does not exist somewhere out there. It is an essential
part of our very existence. This inevitable transition from the quantum
to classical domain taking place every moment of our life represents
the basic premise of the quantum measurement problem. Significantly
in recent times, experimental evidences conspicuously demonstrate
that the distinctive phenomenon like quantum superposition just does
not entirely disappear but inevitably get masked by interactions with
the plethora of particles manifestly operational in the classical
domain.

\pagebreak
In a recent experiment, using a rather sophisticated procedure,
coherent superposition has been demonstrated in a macroscopic object
containing an estimated ten trillion atoms. For this purpose, the
investigators \cite{key-1} used a 40 micron long mechanical resonator,
just large enough to be visible with the naked eye. The resonator
with a resonant frequency of 6.175 GHz to its first excited phonon
state was cooled to a temperature of merely 25 mK over absolute zero
and put in a very high vacuum to minimize environmental effects. Under
these circumstances, the resonator was confirmed to be in its ground
state. Then a signal from a coupled qubit possessing the resonance
frequency of 6.175 GHz was injected into the resonator thereby transferring
the superposition feature of the qubit to the macroscopic object.
Superposition of the ground and the first excited phonon state of
the macroscopic resonator lasted for the resonator relaxation time
of 6.1 ns.

The above demonstration provides strong evidence that quantum mechanics
and its attendant aspect of superposition applies to macroscopic objects
and can be revealed under appropriate circumstances provided that
it was sufficiently decoupled from its environment. Can it apply to
Schr\"{o}dinger's cat? The answer in principle should be yes. But
to prove it, the cat will surely perish for other reasons! Because
in order to conduct the experiment, it would be necessary to remove
all sources of environmental disturbances exposing the cat to exceptionally
low temperatures and high vacuum that would stop the metabolic processes
for its survival. Further examples of coherence of superposition in
large quantum objects have been presented by Bhaumik \cite{key-2}.

More recently, the quantum phenomena essentially arising from quantum
fluctuations and superposition has been demonstrated in an as large
an object as a man size 40 kg mirror in a gravitational detector \cite{key-3}.
The authors succinctly conclude, ``It is remarkable that quantum vacuum
fluctuations can influence the motion of these macroscopic, human-scale
objects, and that the effect is measured.''

These experiments strongly point toward the fact that the quantum
effects of the microscopic world is indeed present in the macroscopic
domain but substantially veiled in their existence by the effects
of some processes for the disappearance of the distinct quantum characteristics
and the appearance of the classical world where we deal with an innumerable
number of particles. Although substantial progress has been made,
exactly how this is accomplished still comprises a subject of an overabundance
of investigations with some intense debates. However, one particular
aspect that is common to all these investigations is the scarcity
of comprehension about where do the probabilities, rather than a certainty,
in quantum measurement come from.

So far, only some \emph{ad hoc} propositions such as Born's rule \cite{key-16}
have allowed the physicists to predict experimental results with uncanny
accuracy of better than a part in trillion \cite{key-12}. But the
basic cause of this essential rule has remained shrouded in a veil
of mystery. One of the prominent investigators in this field, Wojciech
Zurek has attempted to provide a derivation of the Born rule perhaps
to make his program comprehensive \cite{key-4}. But it has faced
a stiff resistance from some foremost investigators including one
of the giants of physics of our time, Nobel laureate Steven Weinberg.

In his classic textbook, \emph{Lectures on Quantum Mechanics}, Weinberg
states \cite[p. 92]{key-5}, ``There seems to be a wide spread impression
that decoherence solves all obstacles to the class of interpretations
of quantum mechanics, which take seriously the dynamical assumptions
of quantum mechanics as applied to everything, including measurement.''
Weinberg goes on to characterize his objection by asserting that the
problem with derivation of the Born's rule by Zurek ``is clearly circular,
because it relies on the formula for expectation values as matrix
elements of operators, which is itself derived from the Born rule.''
In \cite[p. 26]{key-5} he questions, ``If physical states, including
observers and their instruments, evolve deterministically, where do
the probabilities come from? Again in his recent book \cite[p. 131]{key-6},
Weinberg questions, ``So if we regard the whole process of measurement
as being governed by the equations of quantum mechanics, and these
equations are perfectly deterministic, \emph{how do probabilities
get into quantum mechanics}?''

Maximilian Schlosshauer and Arthur Fine remark \cite{key-7}, ``Certainly
Zurek's approach improves our understanding of the probabilistic character
of quantum theory over that sort of proposal by at least one quantum
leap.'' However, they also criticize Zurek's derivation of the Born's
rule of circularity, stating: ``We cannot derive probabilities from
a theory that does not already contain some probabilistic concept;
at some stage, we need to ``put probabilities in to get probabilities
out.''''

In this article, we present a plausible solution of the mysterious
appearance of probabilities from some basic aspects of the well-established
Quantum Field Theory (QFT) of the Standard Model of particle physics.
Our argument relies on some characteristics of the universal quantum
fields that appear to predetermine the values of the complex coefficients
involved in the inherent superposition of eigenstates before measurement.
Thus, it seems the century old mystery of quantum to classical transition
could get a necessary boost from some recently revealed fundamental
properties of the universe through the advent of QFT.

\section{Significant characteristics of quantum fields}

The ultimate ingredient of reality, uncovered by science so far, consists
of fields, which are distinctively non-material in nature. Some perception
of a field can be gained from our daily experience with the classical
field of gravity that pervades us. The field that we do experience
is stable everywhere in our vicinity but varies from place to place
around its origin, the Earth. However, the ultimate realty of quantum
fields also pervade all space including the one in which we exist,
although we have no perception of them what so ever. Unlike the stable
classical fields, however, the quantum fields are distinctly different
in that they are incessantly teeming with intrinsic, spontaneous,
and totally random activity all taking place locally in all space
time elements, from the infinitesimal to the infinite everywhere in
this unimaginably vast universe. Even though, we do not perceive its
lively reality, indisputable evidence of its existence can be found
everywhere in nature with the help of appropriate equipment. An outline
of the salient features of the universal quantum fields are summarized
below:
\begin{itemize}
\item Quantum fields are the primary ingredients of reality, from which
all else is formed, fills all space and time.
\item Every fragment, each spacetime element of the universe, has the same
basic properties as every other fragment.
\item ``The deeper properties of the quantum field theory {[}$\ldots${]}
arise from the need to introduce infinitely many degrees of freedom,
and the possibility that all these degrees of freedom are excited
as quantum mechanical fluctuations.'' \cite[pp. 338-339]{key-8}.
\item Thus the quantum fields are indeed alive with eternal, incessant,
innately spontaneous, totally unpredictable activity of the quantum
fields locally at each space time element even in perfect vacuum at
absolute zero temperature.
\item ``Loosely speaking, energy can be borrowed to make evanescent virtual
particles. Each pair passes away soon after it comes into being, but
new pairs are constantly boiling up, to establish an equilibrium distribution.''
\cite[p. 404]{key-8}
\item One of the most notable aspects of the liveliness of the quantum fields
is the fact that the expectation value or the average value of the
quantum fields has remained immutable almost since the beginning of
time in spite of its unique spontaneous random activities up to infinite
dynamism.
\end{itemize}
Any reasonable concept of physical reality should then owe its eventual
origin to the fundamental reality of quantum fields and their characteristic
attributes. Of particular interest to us is to explore how the incessant,
innately spontaneous and totally unpredictable activity of the primary
reality of quantum fields comprising the overabundance of quantum
fluctuations could foster the probabilistic nature of quantum states.
All fundamental particles are inseparably intertwined in their existence
with the quantum fields. Effects of the quantum fluctuations appear
to have been generally underestimated even though matter would not
have certain exceptional properties like the anomalous $g$ factor
\cite{Schwinger1948} and the Lamb shift \cite{Lamb1947} without
them.

\section{Wave function of an electron}

The elementary particles like electrons, one of the initial products
of material formation from the abstract but physical quantum fields,
are quanta of the fields. However in reality, the physical electron
state is actually a superposition of states produced by the interactions
with the other fields of the standard model. But then the most significant
question is, for a nonrelativistic single electron, what are these
states that are superposed and how do they owe their existence to
the interactions with the other fields? Are these really typical quantum
states or just irregular disturbances in the field?

In order to give a physical depiction of the disturbances of the fields
and quantum fluctuations, quoting from Frank Wilczek \cite[p. 89]{key-10}:
``Here the electromagnetic field gets modified by its interaction
with a spontaneous fluctuation in the electron field---or, in other
words, by its interaction with a virtual electron-positron pair. {[}$\ldots${]}
The virtual pair is a consequence of spontaneous activity in the electron
field. {[}...{]} They lead to complicated, small but very specific
modifications of the force you would calculate from Maxwell's equations.
Those modifications have been observed, precisely, in accurate experiments.''
Emphasizing Wilczek's critical observation again that in spite of
the precipitous transitory characteristics of the virtual particles,
there is an equilibrium distribution \cite[p. 404]{key-8}.

Paraphrasing for further clarification, it turns out that the innately
spontaneous activity of the electron field disturbs the electromagnetic
field around them, and so electrons spend some of their time as a
combination of two disturbances, one in in the electron field and
one in the electromagnetic field. The disturbance in the electron
field is not an electron particle, and the disturbance in the photon
field is not a photon particle. However, the combination of the two
is just such as to be a nice ripple, with a well-defined energy and
momentum, and with an electron's mass. This continues on and on, with
a ripple in any field disturbing, to a greater or lesser degree, all
of the fields with which it directly or even indirectly has an interaction.
So we ascertain that particles are just not simple objects, and although
we often naively describe them as simple ripples in a single field,
that is far from true. Only in a universe with no spontaneous activities---with
no interactions among particles at all---are particles merely ripple
in a single field!

Would not it then be cogent to pronounce that the ``states'' being
superposed here are the irregular disturbances of the fields originating
from the incessant, innately spontaneous, and totally unpredictable
quantum fluctuations? In fact we know quite explicitly what the states
are out of which the physical electron is built, at least order by
order in perturbation theory. The irregular disturbances of the fields
indeed correspond to virtual particles. In particular, their respective
energy-momentum does not correspond to the physical mass of a particle.
One says that these particles are off-shell. However, in the process,
the total energy-momentum is exactly conserved at all times. Because
of the self-interaction of the quantum fields, such an energy-momentum
eigenstate of the field can be expressed as a specific Lorentz covariant
superposition of field shapes of the electron field along with all
the other quantum fields of the Standard Model of particle physics.

It is particularly important to emphasize again that the quantum fluctuations
are transitory but new ones are constantly boiling up to establish
an equilibrium distribution so stable that their contribution to the
screening of the bare charge provided the measured charge of the electron
to be stable up to nine decimal places \cite{Mohr2008} (noteworthy,
the elementary charge is no longer a measurable quantitity because
it is exactly defined since 20 May 2019 by the International System
of Units\cite{key-11}) and the electron $g$-factor results in a
measurement accuracy of better than a part in a trillion \cite{key-12}.

The Lorentz covariant superposition of fluctuations of all the quantum
fields in the one-particle quantum state can be conveniently depicted
leading to a well behaved smooth wave packet. A fairly rigorous underpinning
of the wave packet function for a single particle QFT state in position
space for a scalar quantum field has been provided by Robert Klauber
\cite{key-13}. Since particles of all quantum field are invariably
an admixture of contributions from essentially all the fields of the
Standard Model, the wave packet function of a single particle of a
scalar quantum field can be considered to be qualitatively representative
of those of the spinor and vector quantum fields as well.

Following Klauber, the wave function $\psi(x)$, for an electron in
one dimension, can then be given by the Fourier integral
\begin{equation}
\psi(x)=\frac{1}{\sqrt{2\pi}}\int_{-\infty}^{+\infty}\tilde{\psi}(k)e^{\imath kx}dk\label{eq:1}
\end{equation}
where $\tilde{\psi}(k)$ is a function that quantifies the amount
of each wave number component $k=2\pi/\lambda$ that gets added to
the combination.

From Fourier analysis, we also know that the spatial wave function
$\psi(x)$ and the wave number function $\tilde{\psi}(k)$ are a Fourier
transform pair. Therefore, we can find the wave number function through
the Fourier transform of $\psi(x)$ as
\begin{equation}
\tilde{\psi}(k)=\frac{1}{\sqrt{2\pi}}\int_{-\infty}^{+\infty}\psi(x)e^{-\imath kx}dx
\end{equation}

\begin{figure}[t]
\includegraphics[width=85mm]{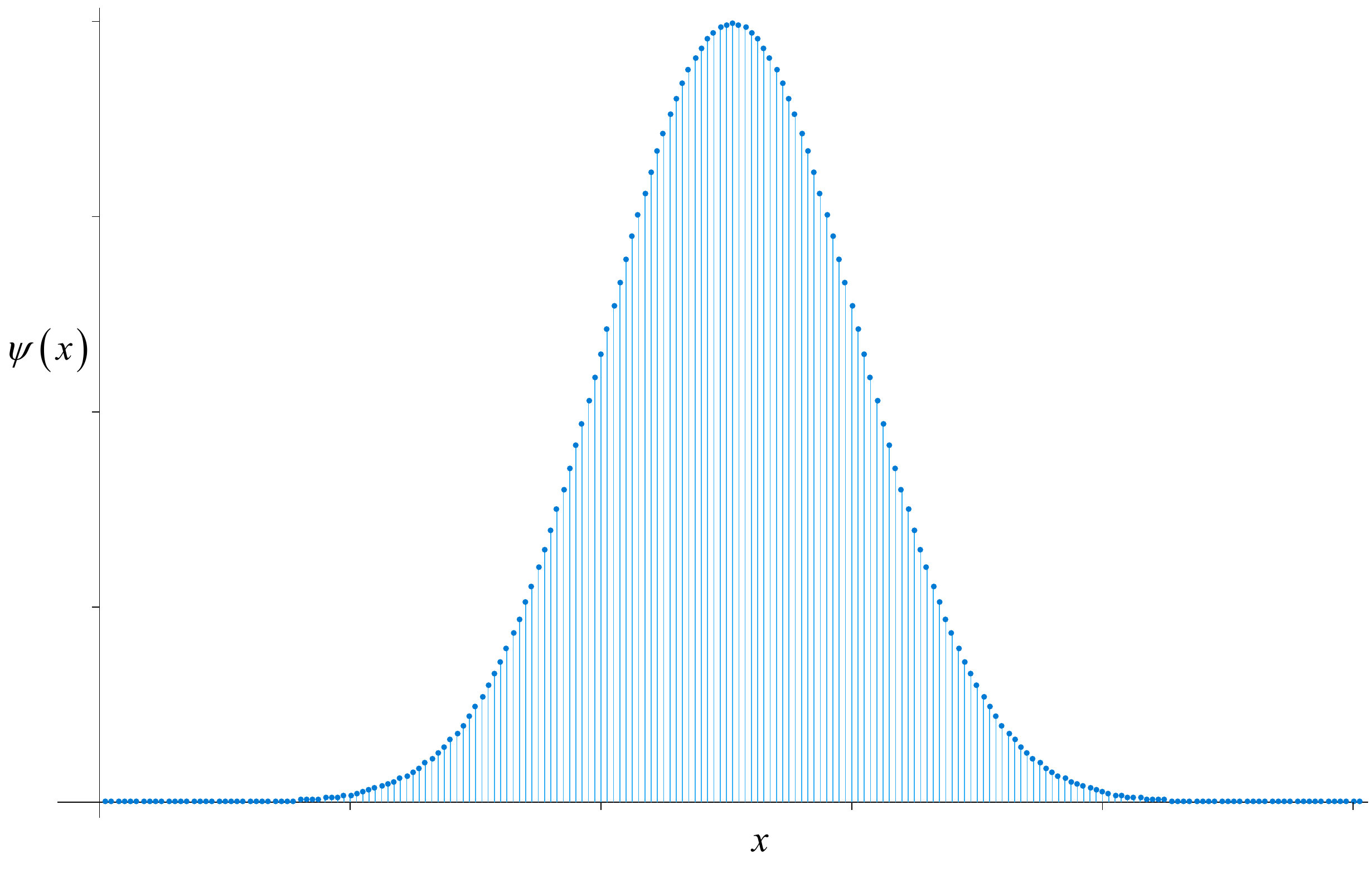}
\caption{\label{fig:1}Plot of Gaussian wave packet function of an electron portrayed
in equation \eqref{eq:1} in position space. The amplitudes represent
genuine reality as they correspond to the contributions from various
quantum fields, the verified ultimate reality uncovered by science so far.}
\end{figure}

Thus the Fourier transform relationship between $\psi(x)$ and $\tilde{\psi}(k)$,
where $x$ and $k$ are known as conjugate variables, can help us
determine the frequency or the wave number content of any spatial
wave function. A plot of the wave function $\psi(x)$ in equation
\eqref{eq:1} gives us the familiar wave function of a quantum particle
like electron (Fig. \ref{fig:1}).

A few unique aspects of this depiction should be noted:
\begin{itemize}
\item First of all, the entire wave function as a whole represents all the
requisite properties of the single electron. Therefore, in measurement,
the entire wave packet should be acquired holistically or nothing
at all. Experimental results demonstrate \cite{key-14} that the entire
extended wave packet can be reduced to the position of measurement
instantaneously quite possibly because of the entanglement of the
wave packet with the wave function of the quantum vacuum.
\item The plot (Fig.~\ref{fig:1}) is a superposition of amplitudes as a
function of position $x$. But what do these amplitudes represent?
As we have repeatedly emphasized, these amplitudes arise from a mixture
of different quantum fields even for a single quantum of the electron
field. Consequently, they have nothing in common except energy, since
the attributes of energy is the same irrespective of which quantum
field they belong to. We are aware that these amplitudes do not constitute
charge distribution of an electron as originally proposed by Schr\"{o}dinger
and shown to be incorrect by Born from his electron scattering experiments.
But then what could these amplitudes mean?
\end{itemize}

\section{Probability amplitudes}

Following Einstein's intuition, it would be cogent to consider that
in measurement, a quantum particle would have the highest probability
of being found where the intensity or the energy density of the quantum
particle is the highest inside the wave packet. Energy density of
a wave is given by the square of its amplitude. Therefore, to get
the probability density, we have to take the square of the amplitude
of the wave function, which usually involves a complex quantity. Consequently,
the absolute square amplitude $\left|\psi(x)\right|^{2}=\psi^{*}(x)\psi(x)$,
which is the probability density function $p(x)$, should represent
the probability density for finding a particle in position space.
Thus
\begin{equation}
p(x)=\left|\psi(x)\right|^{2}
\end{equation}
Since the total probability is 1, the integral
\begin{equation}
\int_{-\infty}^{+\infty}\psi^{*}(x)\psi(x)dx=1
\end{equation}

Max Born did something similar in formulating his famous Born's rule.
Quoting Born from his Nobel Lecture \cite{key-15}: ``Again an idea
of Einstein's gave me the lead. He had tried to make the duality of
particles---light quanta or photons---and waves comprehensible by
interpreting the square of the optical wave amplitudes as probability
density for the occurrence of photons. This concept could at once
be carried over to the $\psi$-function: $\left|\psi(x)\right|^{2}$
ought to represent the probability density for electrons (or other
particles).'' Since then it is known as the Born's rule. However,
Born could not have realized at his time that the wave function of
the electron is derived from real quantum fields and therefore actually
is real and so are the energy density amplitudes that can be described
as probability amplitude, which consequently are real as well. It
appears to be an erudite guess from Born's part, especially judging
from the fact that the single particle wave function was considered
a fictitious mathematical construct at the time and the square of
the wave function was added after submission of the original manuscript
in 1926 \cite{key-16} without any mention of energy density or intensity
involved with wave functions. It is hard to imagine a fictitious mathematical
construct having any energy density or intensity!

By now it should be evident that a quantum particle wave packet function
is indeed real and far from being fictitious. Also Born's rule can
be reasonably derived following Einstein's intuition and does not
need to be a mere postulate. The fact that the position of the electron
is given by a probability instead of certainty should not be surprising
either. It is inevitable since the wave packet function is real. The
Fourier transform correlations between conjugate variable pairs of
any real wave packet have powerful consequences since these variables
obey the uncertainty relation
\begin{equation}
\Delta x\Delta k\geq\frac{1}{2}
\end{equation}
where $\Delta x$ and$\Delta k$ relate to the standard deviations
$\sigma_{x}$ and $\sigma_{k}$ of the wave packet. This is a completely
general property of a wave packet with a reality of its own and is
in fact inherent in the properties of all wave-like systems. It becomes
important in quantum mechanics because of the real wave nature of
particles having the relationship $p=\hbar k$, where $p$ is the
momentum of the particle. Substituting this in the general uncertainty
relationship of a wave packet, the intrinsic uncertainty relation
in quantum mechanics becomes
\begin{equation}
\Delta x\Delta p\geq\frac{1}{2}\hbar
\end{equation}
It is thus evident that a particular fixed value of the position $x$
is not compatible with other measurable quantities like momentum.

However, it is critically important to note that whichever position
$x$ turns up in the measurement process, its probability amplitude
is predetermined from the complex interactions of the various quantum
fields and encoded in the wave packet function. Again emphasizing
from our earlier rather elaborate discussions, it should be aptly
highlighted that the wave function is real and the computable value
of the complex amplitudes $\psi(x)$ in the wave packet function is
preordained from the indispensable interactions of the various quantum
fields and their quantum fluctuations involved in the formation of
the electron wave packet function.

For clarification, the wave number $k$ in the argument of the wave
function for a massive particle like electron obeys the relativistic
relation $p=\hbar k$. Therefore, for electrons $k$ in the exponent
should be replaced by $p/\hbar$.

\section{Probability in Hilbert space}

It is of immense interest to emphasize again that the amplitudes in
the wave packet of a particle resulting from the contributions of
the diverse quantum fields are already predetermined. Hence the probability
distribution of a quantum observable is already preordained before
as well as after the unitary evolution of the Schr\"{o}dinger equation.
Can this comprehension be cogently extended to the observables in
the customary Hilbert space formalism?

In Hilbert space, use of Dirac's abstract algebraic model of bras
and kets, from the bracket notation for the inner product, proved
to be of great computational value. However, there were serious difficulties
in finding a mathematical justification for using them in observables
that have continuous spectrum as in the wave packet function of an
electron we have been exploring so far. These difficulties were circumvented
by the advent of the \emph{rigged Hilbert space} (RHS) in 1960s \cite{Gelfand1964,Berezanskii1968,key-17}.
The RHS is neither an extension nor an interpretation of the physical
principles of Quantum Mechanics, but is simply a mathematical tool
to extract and process the information contained in observables that
have both continuous as well as discrete spectrum. Therefore, in spite
of not using the Dirac's bras and kets for the single particle, the
inferences derived for the single particle wave packet can still be
reasonably extended to the Hilbert space formalism. Observables with
discrete spectrum and a finite number of eigenvectors (e.g., spin)
do not need the RHS. For such observables, the Hilbert space is sufficient
\cite{key-17}.

The Hilbert space is a square integrable, complex, linear, abstract
space of vectors possessing a positive definite inner product assured
to be a number. The states of a quantum mechanical system are vectors
in a multidimensional Hilbert space containing an orthonormal basis
set of eigenfunctions. The observables are Hermitian operators on
that space, and measurements are orthogonal projections. The quantum
wave functions, for example, the solutions of the Schr\"{o}dinger
equation describing physical states in wave mechanics are considered
as the set of components $\psi(x)$ of the abstract vector $\Psi$,
the state vector. However, the state vector does not depend upon any
particular choice of coordinates. The same state vector can be described
in terms of the wave function in position or momentum state or written
as an expansion in wave functions $\psi_{n}$ of definite energy
\begin{equation}
|\Psi\rangle=\sum_{n}\psi_{n}|E_{n}\rangle\label{eq:7}
\end{equation}
suggesting that every linear combination of vectors in a Hilbert space
is again a vector in the Hilbert space. In general, the energy eigenstates
$|E_{n}\rangle$ may not commute with the position eigenstates $|x\rangle$.
The normalized square moduli $\left|\psi_{n}\right|^{2}$ of the energy
complex coefficients are then interpreted as the probability for the
system to be in the energy state $|E_{n}\rangle$ analogous to the
single particle wave function where $\left|\psi(x)\right|^{2}$ is
interpreted as the probability density for the particle to be at $|x\rangle$.

It is worthy to note that the coefficients $\psi_{n}(t)$ change during
unitary time evolution but the probability for measuring outcome in
a given energy eigenstate does not. The solution of the time-dependent
Schr\"{o}dinger equation is given by \cite{key-18}:
\begin{equation}
|\Psi(t)\rangle=\sum_{n}\psi_{n}(0)e^{-\imath\frac{E_{n}}{\hbar}t}|E_{n}\rangle\label{eq:8}
\end{equation}
where $E_{n}$ is the eigenvalue of the corresponding energy eigenvector
$|E_{n}\rangle$. From \eqref{eq:8}, it is seen that for measurements
in the energy basis the time dependence of $\psi_{n}(t)$ during unitary
evolution drops out of the square modulus of the wave vector for computing
the probability 
\begin{equation}
\left|\psi_{n}(0)\right|^{2}e^{-\imath\frac{E_{n}}{\hbar}t}e^{+\imath\frac{E_{n}}{\hbar}t}=\left|\psi_{n}(0)\right|^{2}e^{0}=1\times\left|\psi_{n}(0)\right|^{2}
\end{equation}
and the effective coefficient of superposition remains unchanged.
Measurement in any non-commuting basis, however, leads to quantum
interference effects of the $e^{-\imath\frac{E_{n}}{\hbar}t}$ terms
\cite[pp. 120-123]{Georgiev2017}.

Thus, it seems plausible that the gist of the ideas regarding the
eventual origin of the probabilities from the incessant spontaneous
activities of the ultimate reality of the quantum fields can be extended
to the Hilbert space formalism. After all, Born's rule was first derived
historically for the single quantum particle and subsequently extended
to the Hilbert space.

\subsection{Projective measurement}

Every vector in the Hilbert space, can be expressed in Dirac's notation
as a linear combination \eqref{eq:7} of the energy basis vectors
$|E_{n}\rangle$ with complex coefficients $\psi_{n}$. Multiplying
both sides of \eqref{eq:7} by $\langle E_{m}|$ gives
\begin{equation}
\langle E_{m}|\Psi\rangle=\sum_{n}\psi_{n}\langle E_{m}|E_{n}\rangle
\end{equation}
Since
\begin{equation}
\langle E_{m}|E_{n}\rangle=\delta_{mn}=\begin{cases}
1 & \textrm{if }m=n\\
0 & {\textstyle \textrm{if }}m\neq n
\end{cases}
\end{equation}
it follows that
\begin{equation}
\psi_{n}=\langle E_{n}|\Psi\rangle\label{eq:12}
\end{equation}
which is the transition amplitude of state $|\Psi\rangle$ to state
$|E_{n}\rangle$. The energy basis vectors are then superposition
of quantum states with complex coefficients if viewed in a different
non-commuting basis.

Inserting \eqref{eq:12} into \eqref{eq:7} gives
\begin{equation}
|\Psi\rangle=\sum_{n}|E_{n}\rangle\langle E_{n}|\Psi\rangle\label{eq:13}
\end{equation}
Further defining a projection operator $\hat{P}_{n}=|E_{n}\rangle\langle E_{n}|$
transforms \eqref{eq:13} into
\begin{equation}
|\Psi\rangle=\sum_{n}\hat{P}_{n}|\Psi\rangle
\end{equation}
leading to $\sum_{n}\hat{P}_{n}=\hat{I}$ signifying that the sum
of all the projection operators is unity.

The outer product $|\psi\rangle\langle\psi|$ is called the projection
operator since it projects an input ket vector $|\phi\rangle$ into
a ray defined by the ket $|\psi\rangle$ as follows
\begin{equation}
|\psi\rangle\langle\psi|\,|\phi\rangle=|\psi\rangle\langle\psi|\phi\rangle
\end{equation}
with a probability $\left|\langle\psi|\phi\rangle\right|^{2}$, as
the inner product between two state vectors is a complex number recognized
as the probability amplitude.

This is usually known as projective measurement and we should notice
that it is important for the measurement of a mixed state consisting
of an ensemble of pure states in a density matrix.

\subsection{Operator valued observables}

In a quantum system, what can be measured in an experiment are the
eigenvalues of various observable physical quantities like position,
momentum, energy, etc. These observables are represented by linear,
self-adjoint Hermitian operators acting on Hilbert space.

Each eigenstate of an observable corresponds to eigenvectors $|\psi_{i}\rangle$
of the operator $\hat{A}$, and the associated eigenvalue $\lambda_{i}$
corresponds to the value of the observable in that eigenstate
\begin{equation}
\hat{A}|\psi_{i}\rangle=\lambda_{i}|\psi_{i}\rangle\label{eq:16}
\end{equation}

For a Hermitian operator $\hat{A}$ , the quantum states associated
with different eigenvalues are orthogonal to one another
\begin{equation}
\langle\psi_{i}|\psi_{j}\rangle=\delta_{ij}
\end{equation}

The possible results of a measurement are the eigenvalues of the operator,
which explains the choice of self-adjoint operators for all the eigenvalues
to be real. The probability distribution of an observable in a given
state can be found by computing the spectral decomposition of the
corresponding operator. For a Hermitian operator $\hat{A}$ on an
$n$-dimensional Hilbert space, this can be expressed in terms of
its eigenvalues following \eqref{eq:16} as
\begin{equation}
\hat{A}=\sum_{i}\lambda_{i}|\psi_{i}\rangle\langle\psi_{i}|
\end{equation}

If the observable $\hat{A}$, with eigenstates $\left\{ |\psi_{i}\rangle\right\} $
and spectrum $\left\{ \lambda_{i}\right\} $ is measured on a system
described by the state vector $|\Psi\rangle$, the probability for
the measurement to yield the value $\lambda_{i}$ is given by
\begin{equation}
p\left(\lambda_{i}\right)=\left|\langle\psi_{i}|\Psi\rangle\right|^{2}
\end{equation}
This again is the famous Born's rule and we can see that it can be
derived by extending the concepts discussed in the case of the single
particle. After the measurement the system is in the eigenstate $|\psi_{i}\rangle$
corresponding to the eigenvalue $\lambda_{i}$ found in the measurement,
which is called the reduction of state.

Recalling the discussions of the probability amplitudes in the one
particle wave function, the probability amplitudes of the quantum
states involved in superposition in Hilbert space has likewise been
predetermined very possibly again because of the characteristic ceaseless
activity and mutual interactions of the quantum fields. A quantum
state in superposition generally has non-zero values for all states
in superposition \cite{key-6}. This is why we assert that the energy
of the states is not on the mass shell. Again, this could be possible
because of the interactions of the quantum states in superposition
with the ceaseless quantum fluctuations just as in the case for the
superposed components making up the composition of the structure of
the non relativistic single electron.

It is of paramount importance to reemphasize that matter and particularly
the elementary particles comprising them would not have some behavior
in the absence of the special characteristics of the quantum fields
listed earlier. These distinct activities are well recognized in the
Lamb shift, anomalous electron $g$-factor, etc. Quantum superposition
with definite complex amplitudes can therefore be also an example
of such behavior.

\subsection{Stern--Gerlach experiment}

The Stern--Gerlach experiment is the most striking illustration of
the experimental implementation of quantum measurements. It is as
simple as it is persuasive. The following Stern--Gerlach experiment
carried out using neutrons, each having a spin $\frac{1}{2}$, reinforces
the fact that the probabilities of the eigenstates in superposition
is present from the beginning, very likely because of the incessant
activities of the quantum fields. The results of a series of Stern--Gerlach
setups in tandem \cite{key-19} show that the precise probabilities
in superposition are indeed restored repeatedly following each projective
measurement. After examining the results of these simple experiments,
it is hard not be convinced about our assertion that the coefficients
of probabilities are preexistent in superposition of states and their
effect in the measurement of probability do not change during the
unitary evolution of the superposed system.

\begin{figure*}[t]
\includegraphics[width=170mm]{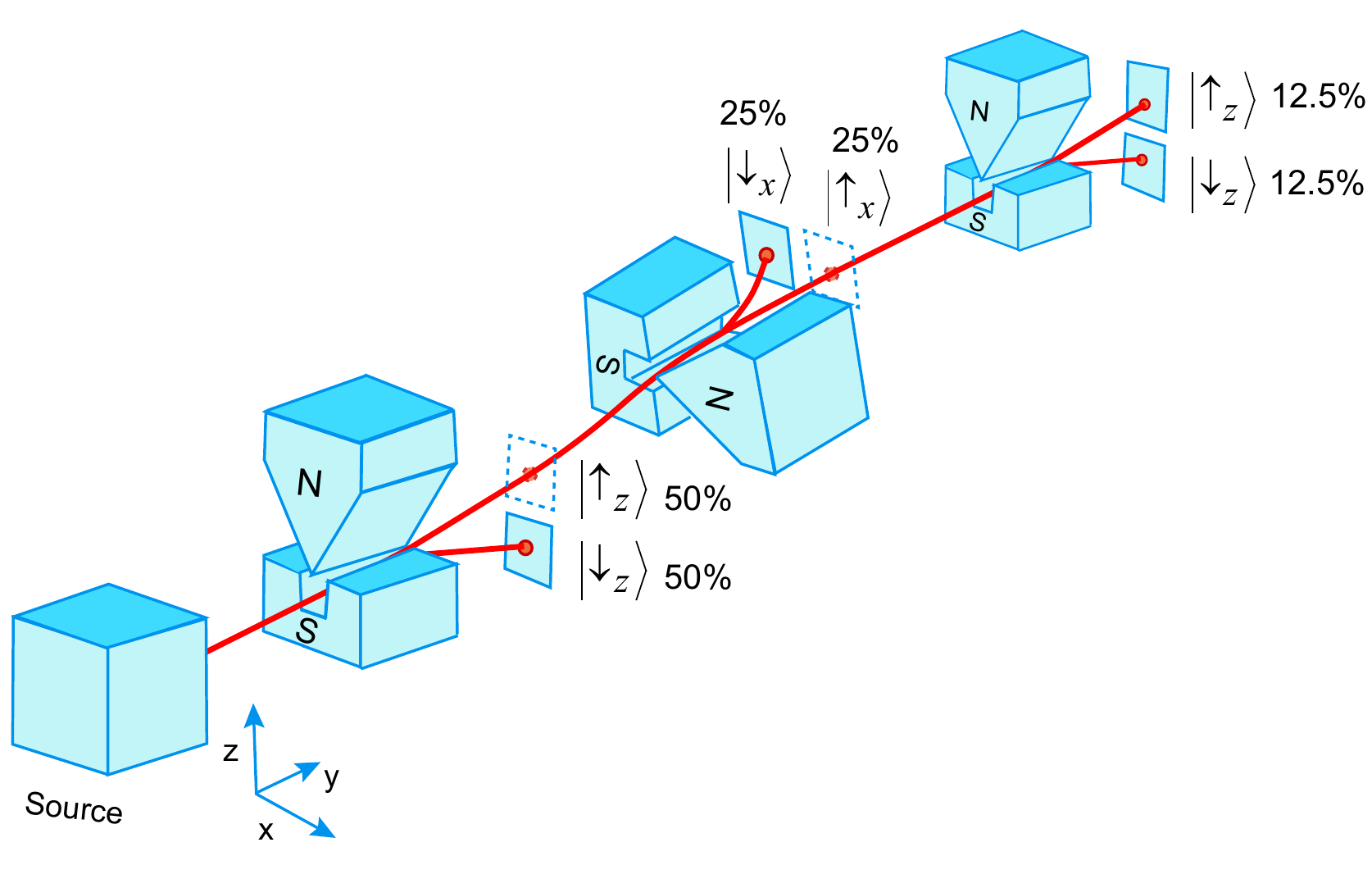}
\caption{\label{fig:2} Serial Stern--Gerlach experiment in which neutrons are fired from a source into the non-homogeneous magnetic fields of three sequential Stern--Gerlach magnets oriented along the $z$-axis, the $x$-axis and the $z$-axis, respectively.}
\end{figure*}

For the purpose of presentation clarity, we will assume that before entering the first Stern--Gerlach magnet (Fig.~\ref{fig:2}), the direction of the neutron
spin magnetic moment is in a definite superposed state of two states referred to as spin-up and spin-down
\begin{equation}
|\Psi_{0}\rangle= \alpha |\uparrow_{z}\rangle+ \beta |\downarrow_{z}\rangle \label{eq:20}
\end{equation}
with unknown complex coefficients $\alpha$ and $\beta$ that are constrained by
$|\alpha|^2=|\beta|^2=\frac{1}{2}$.

The superposed state \eqref{eq:20} reduces (or collapses) as soon as
the neutron enters the magnets of the analyzer to just one
spin-$z$ direction by immediate momentum and energy transfer with the
magnet, rather than by subsequent determination at the screen \cite{key-20}.
After exiting the magnet aligned in the $z$ direction, the trajectory
of the neutron spin that can take only two equal but opposite values,
will be deflected in either the $z+$ or $z-$ directions. If we
denote these states by $|\uparrow_{z}\rangle$ and $|\downarrow_{z}\rangle$
respectively, we could say that the initial state performs one out of two
equally probable quantum jumps
\begin{equation}
\begin{cases}
|\Psi_{0}\rangle\to|\uparrow_{z}\rangle & \textrm{with }p=\frac{1}{2}\\
|\Psi_{0}\rangle\to|\downarrow_{z}\rangle & \textrm{with }p=\frac{1}{2}
\end{cases}
\end{equation}
When the beam of neutrons hits a detector screen, two
spatially separated spots will appear corresponding to the two distinct
trajectories. Each of the two spots would show equal number of neutrons
following $|\alpha|^2=|\beta|^2=\frac{1}{2}$. If we now choose to send only the $|\uparrow_{z}\rangle$
state through second Stern--Gerlach magnet aligned in the $z$ direction,
all the neutrons will be found, consistent with its preparation, in
the upper region only.

However, if the state \mbox{$|\uparrow_{z}\rangle=\frac{1}{\sqrt{2}}\left(|\uparrow_{x}\rangle+|\downarrow_{x}\rangle\right)$}
faces Stern--Gerlach magnet aligned along the noncommuting orthogonal
$x$-axis, any previous information about $|\uparrow_{z}\rangle$
will be completely destroyed and the direction of the spin magnetic
moment will no longer be in an eigenstate of~$\hat{\sigma}_z$ due to occurrence of one out of two
equally probable quantum jumps
\begin{equation}
\begin{cases}
|\uparrow_{z}\rangle\to|\uparrow_{x}\rangle & \textrm{with }p=\frac{1}{2}\\
|\uparrow_{z}\rangle\to|\downarrow_{x}\rangle & \textrm{with }p=\frac{1}{2}
\end{cases}
\end{equation}
Therefore, if only the $z+$ neutrons are passed through a second
Stern--Gerlach magnet, which measures the neutron's $x$-spins, the
neutrons are deflected either right or left, labeled $x+$ and $x-$,
and the number of neutrons with $|\uparrow_{x}\rangle$ and $|\downarrow_{x}\rangle$
spin is split even as expected.

Subsequently, if we pass only the \mbox{$|\uparrow_{x}\rangle$} neutrons
through a third Stern--Gerlach magnet oriented along the orthogonal
$z$-direction, we observe that their previous \mbox{$z$-spin} value has been reset,
and they are again split evenly between $z+$ and $z-$. This is despite
the fact that we selected only the $z+$ neutrons from the first Stern--Gerlach
magnet. When the second one is measured, it resets the state of the
first one. Thus, there is another clear indication that the complex
coefficients of superposition are predetermined, again very possibly
by the quantum fluctuations. We can hence infer that the ceaseless
quantum fluctuations as well as the mutual interactions of the quantum
fields preordain the probability of detection of a quantum state.

In a recent investigation \cite{key-21}, by performing an ingenious
experiment involving superposition of three eigenstates of a state
vector given by
\begin{equation}
|\Psi\rangle=\alpha_{0}|0\rangle+\alpha_{1}|1\rangle+\alpha_{2}|2\rangle
\end{equation}
it was strikingly demonstrated that the complex coefficient $\alpha_{0}$
governing the probability of the particular quantum state $|0\rangle$
in a superposition of three states can be measured without affecting
the superposition of the two other remaining states in superposition.
This further reinforces the fact that the coefficients of superposition
that determines the probability outcome of measurement are predetermined.

\section{Conclusion}

History of the development of quantum mechanics is replete with a
notable trend. Because of the sheer novelty of the subject so remarkably
different from the established classical physics, the pioneers of
the development of quantum physics utilized a procedure quite frequently
with notable success. Due to a thorough lack of experience with the
precepts of the newly emerging subject of quantum mechanics, an empirical
model was fashioned first to accommodate the observed information.
The experiential model was then amended to accommodate a more realistic
version from a deeper understanding gained from subsequent revelations.
This successful procedure started almost from the beginning with the
proposal of a quantum by Max Planck.

Out of sheer frustration of not being able to match the characteristics
of blackbody radiation to his equation, Planck introduced the indivisible
radiation quantum believing it was just a necessary mathematical oddity
without having any reality whatsoever. Five years later, Einstein
persuasively demonstrated the reality of the quantum from the results
of photo-electric effect. Citing another example, Neils Bohr crafted
the first atomic model with discrete electron orbits merely to fit
the observed spectral data. With the proposal and subsequent verification
of matter wave, Schr\"{o}dinger demonstrated Bohr's discrete atomic
orbits to be real standing wave patterns of matter wave and the list
continues.

A similar situation presented itself with the fabrication of the wave
packet function to accommodate the observed wave particle duality.
It has essentially been considered to be a rather fictitious mathematical
construct giving the probability density amplitude following Max Born's
educated guess. From our current deeper understanding based on contemporary
knowledge of the primary reality of the quantum fields and their incessant,
innately spontaneous, totally unpredictable activities identified
as quantum fluctuations, we now realize that the wave packet function
for a single nonrelativistic electron is in fact real. The wave function
represents among others, real energy density amplitudes of the electron
and consequently the probability density amplitude following Einstein's
intuition. Since the amplitudes of the wave function in position space
are computable, the probability of the electron at a particular position
is predetermined as a distinct result of the specific activities of
the quantum fields. This phenomenon can be cogently extended to the
superposition of quantum states in Hilbert space.

We, therefore believe that a plausible answer has now been provided
to the question, where do the probabilities in the measurement come
from, thus removing one of the hurdles in resolving the century old
measurement problem.

\section*{Acknowledgments}

The author wishes to acknowledge helpful discussions with Zvi Bern,
Eric D'Hoker and Danko Georgiev.

\balance


\begin{thebibliography}{10}
\expandafter\ifx\csname url\endcsname\relax
  \def\url#1{\texttt{#1}}\fi
\expandafter\ifx\csname urlprefix\endcsname\relax\def\urlprefix{URL }\fi
\expandafter\ifx\csname href\endcsname\relax
  \def\href#1#2{#2} \def\path#1{#1}\fi

\bibitem{Kross-2021}
B.~Kross. How many atoms are in the human body?. \emph{Thomas Jefferson
  National Accelerator Facility - Office of Science Education} 2021;
  \url{https://education.jlab.org/qa/mathatom_04.html}.

\bibitem{key-1}
A.~D. O'Connell, M.~Hofheinz, M.~Ansmann, R.~C. Bialczak, M.~Lenander,
  E.~Lucero, M.~Neeley, D.~Sank, H.~Wang, M.~Weides, J.~Wenner, J.~M. Martinis,
  A.~N. Cleland. Quantum ground state and single-phonon control of a mechanical
  resonator. \emph{Nature} 2010; \textbf{464}(7289):697--703.
\newblock \href {http://doi.org/10.1038/nature08967}
  {\path{doi:10.1038/nature08967}}.

\bibitem{key-2}
M.~L. Bhaumik. Is {S}chr\"{o}dinger's cat alive?. \emph{Quanta} 2017;
  \textbf{6}:70--80.
\newblock \href {http://doi.org/10.12743/quanta.v6i1.68}
  {\path{doi:10.12743/quanta.v6i1.68}}.

\bibitem{key-3}
H.~Yu, L.~McCuller, M.~Tse, N.~Kijbunchoo, L.~Barsotti, N.~Mavalvala, et~al..
  Quantum correlations between light and the kilogram-mass mirrors of {LIGO}.
  \emph{Nature} 2020; \textbf{583}(7814):43--47.
\newblock \href {http://doi.org/10.1038/s41586-020-2420-8}
  {\path{doi:10.1038/s41586-020-2420-8}}.

\bibitem{key-16}
M.~Born. Zur {Q}uantenmechanik der {S}to{\ss}vorg\"{a}nge. \emph{Zeitschrift
  f\"{u}r Physik} 1926; \textbf{37}(12):863--867.
\newblock \href {http://doi.org/10.1007/bf01397477}
  {\path{doi:10.1007/bf01397477}}.

\bibitem{key-12}
B.~Odom, D.~Hanneke, B.~D'Urso, G.~Gabrielse. New measurement of the electron
  magnetic moment using a one-electron quantum cyclotron. \emph{Physical Review
  Letters} 2006; \textbf{97}(3):030801.
\newblock \href {http://doi.org/10.1103/PhysRevLett.97.030801}
  {\path{doi:10.1103/PhysRevLett.97.030801}}.

\bibitem{key-4}
W.~H. Zurek. Environment-assisted invariance, entanglement, and probabilities
  in quantum physics. \emph{Physical Review Letters} 2003;
  \textbf{90}(12):120404.
\newblock \href {http://doi.org/10.1103/PhysRevLett.90.120404}
  {\path{doi:10.1103/PhysRevLett.90.120404}}.

\bibitem{key-5}
S.~Weinberg. Lectures on Quantum Mechanics. Cambridge University Press,
  Cambridge, 2013.

\bibitem{key-6}
S.~Weinberg. Third Thoughts. Harvard University Press, Cambridge,
  Massachusetts, 2018.

\bibitem{key-7}
M.~Schlosshauer, A.~Fine. On {Z}urek's derivation of the {B}orn rule.
  \emph{Foundations of Physics} 2005; \textbf{35}(2):197--213.
\newblock \href {http://doi.org/10.1007/s10701-004-1941-6}
  {\path{doi:10.1007/s10701-004-1941-6}}.

\bibitem{key-8}
F.~Wilczek. Fantastic Realities: 49 Mind Journeys and A Trip to Stockholm.
  World Scientific, Singapore, 2006.
\newblock \href {http://doi.org/10.1142/6019} {\path{doi:10.1142/6019}}.

\bibitem{Schwinger1948}
J.~Schwinger. On quantum-electrodynamics and the magnetic moment of the
  electron. \emph{Physical Review} 1948; \textbf{73}(4):416--417.
\newblock \href {http://doi.org/10.1103/PhysRev.73.416}
  {\path{doi:10.1103/PhysRev.73.416}}.

\bibitem{Lamb1947}
W.~E. Lamb, R.~C. Retherford. Fine structure of the hydrogen atom by a
  microwave method. \emph{Physical Review} 1947; \textbf{72}(3):241--243.
\newblock \href {http://doi.org/10.1103/PhysRev.72.241}
  {\path{doi:10.1103/PhysRev.72.241}}.

\bibitem{key-10}
F.~Wilczek. The Lightness of Being: Mass, Ether, and the Unification of Forces.
  Basic Books, New York, 2008.

\bibitem{Mohr2008}
P.~J. Mohr, B.~N. Taylor, D.~B. Newell. {CODATA} recommended values of the
  fundamental physical constants: 2006. \emph{Reviews of Modern Physics} 2008;
  \textbf{80}(2):633--730.
\newblock \href {http://doi.org/10.1103/RevModPhys.80.633}
  {\path{doi:10.1103/RevModPhys.80.633}}.

\bibitem{key-11}
D.~B. Newell, E.~Tiesinga. The International System of Units (SI). NIST Special
  Publication 330. National Institute of Standards and Technology,
  Gaithersburg, Maryland, 2019.
\newblock \href {http://doi.org/10.6028/nist.sp.330-2019}
  {\path{doi:10.6028/nist.sp.330-2019}}.

\bibitem{key-13}
R.~D. Klauber. Student Friendly Quantum Field Theory. 2nd Edition. Sandtrove
  Press, Fairfield, Iowa, 2015.

\bibitem{key-14}
M.~Fuwa, S.~Takeda, M.~Zwierz, H.~M. Wiseman, A.~Furusawa. Experimental proof
  of nonlocal wavefunction collapse for a single particle using homodyne
  measurements. \emph{Nature Communications} 2015; \textbf{6}:6665.
\newblock \href {http://doi.org/10.1038/ncomms7665}
  {\path{doi:10.1038/ncomms7665}}.

\bibitem{key-15}
M.~Born. Statistical interpretation of quantum mechanics. \emph{Science} 1955;
  \textbf{122}(3172):675--679.
\newblock \href {http://doi.org/10.1126/science.122.3172.675}
  {\path{doi:10.1126/science.122.3172.675}}.

\bibitem{Gelfand1964}
I.~M. Gelfand, N.~Y. Vilenkin. Generalized Functions, Volume 4: Applications of
  Harmonic Analysis. Academic Press, New York, 1964.

\bibitem{Berezanskii1968}
Y.~M. Berezanskii. Expansions in Eigenfunctions of Selfadjoint Operators.
  Vol.~17 of \emph{Translations of Mathematical Monographs}. American
  Mathematical Society, 1968.

\bibitem{key-17}
R.~de~la Madrid. The role of the rigged {H}ilbert space in quantum mechanics.
  \emph{European Journal of Physics} 2005; \textbf{26}(2):287--312.
\newblock \href {http://arxiv.org/abs/quant-ph/0502053}
  {\path{arXiv:quant-ph/0502053}}. \href
  {http://doi.org/10.1088/0143-0807/26/2/008}
  {\path{doi:10.1088/0143-0807/26/2/008}}.

\bibitem{key-18}
L.~Susskind, A.~Friedman. Quantum Mechanics: The Theoretical Minimum. What You
  Need to Know to Start Doing Physics. Basic Books, New York, 2014.

\bibitem{Georgiev2017}
D.~D. Georgiev. Quantum Information and Consciousness: A Gentle Introduction.
  CRC Press, Boca Raton, 2017.
\newblock \href {http://doi.org/10.1201/9780203732519}
  {\path{doi:10.1201/9780203732519}}.

\bibitem{key-19}
J.~J. Sakurai, J.~J. Napolitano. Modern Quantum Mechanics. Cambridge University
  Press, Cambridge, 2020.
\newblock \href {http://doi.org/10.1017/9781108587280}
  {\path{doi:10.1017/9781108587280}}.

\bibitem{key-20}
M.~Devereux. Reduction of the atomic wavefunction in the {S}tern--{G}erlach
  magnetic field. \emph{Canadian Journal of Physics} 2015;
  \textbf{93}(11):1382--1390.
\newblock \href {http://doi.org/10.1139/cjp-2015-0031}
  {\path{doi:10.1139/cjp-2015-0031}}.

\bibitem{key-21}
F.~Pokorny, C.~Zhang, G.~Higgins, A.~Cabello, M.~Kleinmann, M.~Hennrich.
  Tracking the dynamics of an ideal quantum measurement. \emph{Physical Review
  Letters} 2020; \textbf{124}(8):080401.
\newblock \href {http://doi.org/10.1103/PhysRevLett.124.080401}
  {\path{doi:10.1103/PhysRevLett.124.080401}}.

\end{thebibliography}
\end{document}